# Entirely Transformerless Universal Direct Injection Power-Flow Controller

Davood Keshavarzi, Alexander Koehler, Wolfram H. Wellssow, and Stefan M. Goetz

*Abstract*—An increasing penetration of renewable energy resources, electric vehicle chargers, and energy storage systems into low-voltage power grids causes several power management and stability problems, such as reverse power flow, (local) overload lines, and over- / under-voltage. Previous power-flow and soft-open-point solutions are bulky and expensive. They need transformers and large magnetics, some on grid frequency, others more compact at high frequency. Even suggested circuits with high-frequency transformers still struggle with cost and size. We present a compact partial power-conversion high-current full-power-flow control circuit without a single transformer. We combine silicon and silicon-carbide, each with their specific advantages for current-dense direct injection. The circuit further needs fewer semiconductors than previous concepts. The circuit links a shunt converter through a non-isolated inverter bidirectionally with low-voltage series modules that practically float with their respective phases can serve between different feeders in low-voltage power grids. We analyze the circuit mathematically and evaluate the operation in simulation and experimental results.

## I. Introduction

THE structure of power generation in electrical grids is undergoing a fundamental shift. The decommissioning of coal-fired power plants for environmental reasons and the continuous expansion of renewable energy sources are pushing the electrical power grid—particularly low-voltage (LV) grid—closer to its limits. The widespread integration of roof-top photovoltaic units, electric vehicle charger stations, and energy storage systems with intermittent behavior changes reliability and power quality in LV and last-mile distribution grids [1]. The massive solar integration introduces bidirectional power flows, sometimes multiple times the load for which the LV grids were originally designed [2]. Consequently, the local grid voltage level fluctuates and can simultaneously violate both the upper and lower limit of ±10% deviation (European Standard EN50160) even in a single feeder [3]. Solutions to these problems are pressing to ensure a high level of supply reliability and quality.

Various solutions promised to solve these problems. Grid expansion, which replaces existing cables with higher-capacity alternatives or installs additional lines, is often associated with high costs and extensive underground construction efforts [4, 5].

On-load tap changers, voltage regulators, switch shunt capacitors, and reactive power management systems are effective for increasing or decreasing the voltage across an entire feeder or even all feeders of a grid section [6-8]. In many locations, however, the conditions are not as homogeneous and the voltage varies across feeders and even within feeders. Traditional installations cannot solve such local voltage and power-flow problems effectively. Beyond voltage stabilization, it does not resolve the issue of thermal overload. Thus, they leave a critical aspect of grid reliability untended.

Unified power flow controllers (UPFCs) promise targeted load flow management but are typically bulky and require significant space; accommodation in cable distribution cabinets appears illusionary [9, 10]. Additionally, existing power-flow controllers are either limited in their capacity or entirely incapable of compensating harmonics and voltage distortions.

A promising application for power quality enhancement and flow control in LV grids is the use of soft open points (SOPs) [12, 13]. SOPs are power electronic systems typically installed at normally open points in distribution grids to enable precise voltage regulation and power flow control [14]. Unlike UPFCs, most SOP designs rely on back-to-back converters, where power is transferred between feeders via an AC-to-DC-to-AC process. While effective, these converters must handle the full line voltage and current directly with their transistors, which can result in higher costs and energy losses.

Recent advancements in semiconductor technology, including wide-bandgap materials and transistor-cell miniaturization in silicon, have opened up new opportunities for power electronics. These advancements enable the direct grid connection of devices with high voltage and current ratings. Additionally, novel circuit topologies have emerged that employ partial-power processing to minimize size and reduce power loss by converting only a fraction of the total power flow.

Lu et al. previously introduced a direct-injection concept for SOP and UPFC systems, which employs shunt and series converters directly tied to the grid, thereby eliminating the need for bulky transformers [15]. Similarly, a hybrid approach using gallium nitride (GaN) devices was proposed to avoid the use of large, discrete high-current filtering inductors [16]. However, both of these approaches rely on three full-size isolated bidirectional DC-DC converters, one for each floating module, which significantly increases the overall cost and footprint of the system.

This paper proposes a compact transformerless power-flow controller with an optimized control strategy. The circuit eliminates all transformers, also high-frequency ones. Instead, the circuit directly connects the fully electronic shunt and series stages to the grid. Therefore, the proposed topology does not require isolated DC-DC converters for each phase anymore. The central component of the proposed power-flow controller is an interconnecting H-bridge converter for each phase, which facilitates power sharing between shunt and series stages as well as floating dc-link voltage for each series modules.

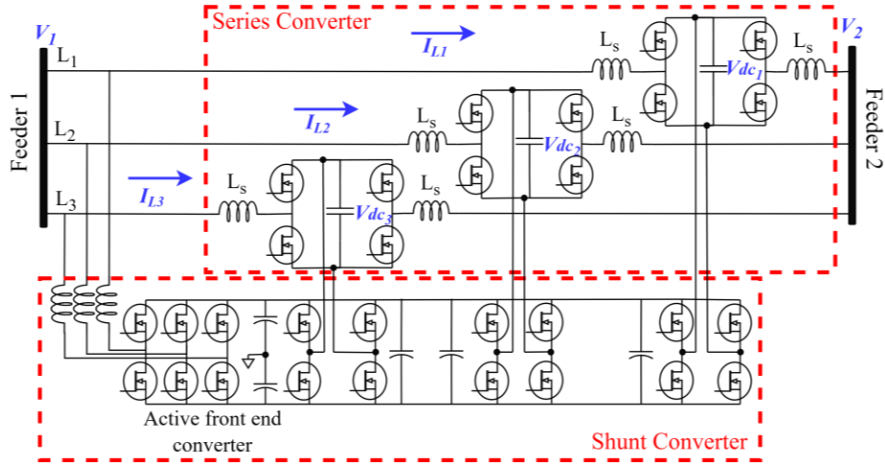

Fig. 1. Circuit diagram of the proposed power-flow controller.

Beyond its voltage regulation capabilities, the proposed controller allows for power-flow control in LV grids without the need for remote throttling of distributed power generation or grid reconfiguration. In looped or segmented grid areas supplied by multiple feeders, such as SOPs, the controller can adjust power flow to prevent thermal overloads, improve stability margins, and meet contractual requirements without the need for power restrictions. Since the circuit can use higher-level control and strategies from SOPs and UPFCs, the paper focuses on the topology. The main contributions are as follows:
- In contrast to back-to-back converters, in which semiconductors need to meet full voltage and current rating, the proposed circuit employs high-voltage low-current components in the shunt stage and low-voltage high-current semiconductors in the series stage.
- The proposed circuit only needs to convert a fraction of the (re-)injected power whereas back-to-back converters must handle full line power.
- Compared to conventional UPFCs, which typically incorporate bulky and bandwidth-limited series (or shunt) transformer(s), the proposed power-flow controller eliminates them, which reduces the overall cost and size of the system.
- Compared to other universal direct-injection power-flow controllers, the proposed circuit can avoid any transformer, even high-frequency links, which are still sizable. Instead, it links the shunt and series stage with a simple nonisolated interconnecting H-bridges.
- Our solution further saves semiconductors compared to earlier concepts with bidirectional isolated DC-DC stages.

In the following, Section II discusses system configuration,
modeling, and control strategy for stable operation. Section III designs the control of the H-bridges. The control of active and reactive power follows in Section IV. Section V presents an analysis of the simulation results, and experimental validation is provided in Section VI. Finally, Section VII concludes the paper with a summary of the findings.

## II. System configuration

Fig. 1 shows the overall configuration of the presented circuit. The series-injection modules are placed in series with grid lines and inject arbitrary current bidirectionally into the grid. Since the line impedance is small in distribution systems, a small differential voltage (typically < 15% of nominal voltage) can drive large current flows. The shunt-injection stage connects directly to the grid lines and provides the desired low-voltage dc-link for each series modules floating with dedicated phase voltage. Since the series module can inject/sink power to/from the grid, the shunt stage must have bidirectional functionality. Therefore, it consists of an active-front-end converter and three interconnecting H-bridges, which interface the series modules directly. This circuit can be considered as partial-power design full-power-flow functionality controller. It only needs to convert a small share to control a large power flow. In the following, each stage is discussed in more detail. Since it is connected in series with the grid lines and the dc-bus voltage is low (about 50 V), high-current low-voltage Si MOSFET technology can handle not only the rated current but also short-circuit current. This feature makes the circuit unique and practicable.

### A. Series-injection stage

The series-injection stage plays the key role in power-flow control by regulating series ac voltage with adjustable magnitude and phase angle. It consists of an H-bridge module, a dc-link capacitor, and series inductors on both sides for each phase. Fig. 2 (a) represents the single-phase circuit diagram of the series-injection stage placed between two ordinary feeders of the LV grid. The line current follows

$$I_L = \frac{V_1 + V_S - V_2}{Z_1 + Z_2},\quad (1)$$

where $V_1$ and $V_2$ are feeders' open-circuit voltage, $V_S$ is the series module's voltage, and $Z_1$ and $Z_2$ are lines impedances. Accordingly, the line current can be controlled by regulating the series module's voltage. It is assumed that the dc-link voltage is constant, the series-injection stage can be modeled as a linear time-invariant system. For each phase, we can derive a current equation in the d-q rotating frame as

$$V_{1,d} = V_{2,d} - V_{SM,d} - \omega_g L_s I_q,$$
$$V_{1,d} = V_{2,d} - V_{SM,d} + \omega_g L_s I_d, \quad (2)$$

where $\omega_g$ and $L_s$ are grid frequency and the lines' total inductivities [17]. Linear PI controllers can regulate the current in the d-q frame. Fig. 2 (b) illustrates the control block diagram of the series-injection module for each phase. The controller has a single loop that injects synchronized current into the grid lines, independently.

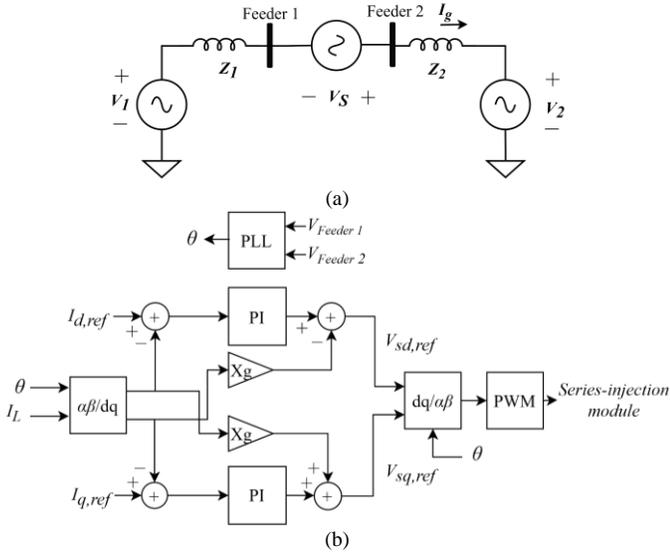

Fig. 2. Series-injection stage: (a) single diagram of module between two grid feeders, (b) controller block diagram of series-injection module.

### B. Shunt-injection stage

Instead of any isolated dc-dc converters as in previous systems, we propose an active-front-end (AFE) converter in combination with three H-bridge modules to supply the injection modules and serve as shunt compensator (static compensator) to the line. The AFE converter acts as an interconnecting bridge between the ac mains and dc side. It offers bidirectional power flow, power-factor control, and reactive power as well as harmonics absorption [18]. Fig. 3(a) shows the circuit diagram of the AFE converter. It consists of a three-leg four-wire SiC MOSFET inverter with split dc-bus connected to the grid lines through an LC-filter. We present a voltage-oriented control with PI controller for AFE converter [19]. It reduces the steady-state error compared to other controllers [20]. The AFE currents are transformed into the d-q rotating frame synchronous to the filter capacitor's voltage for control robustness to grid disturbances. The AFE's voltage follows

$$U_d = V_{C,d} - K_P \left(1 + \frac{1}{\tau_i s}\right)\left(I_d^{ref} - I_d\right) + \omega_g L_f I_q,$$
$$U_q = V_{C,q} - K_P \left(1 + \frac{1}{\tau_i s}\right)\left(I_q^{ref} - I_q\right) - \omega_g L_f I_d. \quad (3)$$

Fig. 3(b) represents the control diagram of the AFE converter, which includes outer and inner control loops to control dc-bus voltage and reactive power at the ac side. The dc-bus voltage of the AFE converter should be high enough (i.e., ≥ 700 V) to suppress current distortion [21, 22]. The proportional gain ($K_P$) and the integral constant ($\tau_i$) can be obtained as

$$K_P = \frac{L_f}{1.5 a T_s},$$
$$a = \frac{1}{\frac{\pi}{2} - \varphi_m}, \quad (4)$$
$$\tau_i = \frac{L_f}{R_f},$$
$$T_s = \text{sampling time},$$

where $\varphi_m$ is the phase margin of the controller and $R_f$ the inductor resistance [23]. In control theory, a phase margin greater than 45 degrees is usually suggested for stable operation. Therefore, $a > 2$ is chosen as stability criterion.

It is obvious that the AFE's output voltage cannot be connected to the series-injection modules directly. The injection modules practically float up and down with respective line's potential. The three interconnecting H-bridge converters are responsible for regulating the low-voltage dc-link of series floating modules. Fig. 4 presents the detailed circuit configuration of the interconnecting H-bridge modules.

Each module includes a two-leg inverter, dc-link capacitor, and common- / differential-mode filters. Each leg of the H-bridge converter is modulated with the grid voltage but offset to generate half of the module dc voltage. Therefore, the differential voltage dictates the dc-link voltage while common mode current must be compensated by regulating common mode voltage.

### III. CONTROL OF ACTIVE AND REACTIVE POWER

The series-injection module's voltage determines the power flow through grid lines. Like other power electronics systems, the proposed power-flow controller has an operating area that guarantees safe operation. The series module's voltage can be written as

$$V_S = r\vec{V_1}e^{j\gamma}, \quad (5)$$

where $0 < r < r_{max}$ and $0 < \gamma < 2\pi$. Without over-modulation, $r_{max}$ is equal to $\frac{V_{dc}}{\sqrt{2}V_1}$.

Installed between two feeders, the injected power into Feeder 2 agrees to

$$S_{inj} = V_2 I_g^* =$$
$$V_2 \angle \theta_2 \left[ \frac{V_1 \angle \theta_1 + rV_1 \angle \theta_1 + \gamma}{X_g \angle 90} \right]^* =$$
$$\begin{cases} P_{inj} = \frac{V_1 V_2}{X_g} \sin(\theta_1 - \theta_2) + \frac{rV_1 V_2}{X_g} \sin(\theta_1 - \theta_2 + \gamma) \\ Q_{inj} = \frac{V_1 V_2}{X_g} \cos(\theta_1 - \theta_2) + \frac{rV_1 V_2}{X_g} \cos(\theta_1 - \theta_2 +) \\ \qquad - \frac{V_2^2}{X_g}. \end{cases} \quad (6)$$

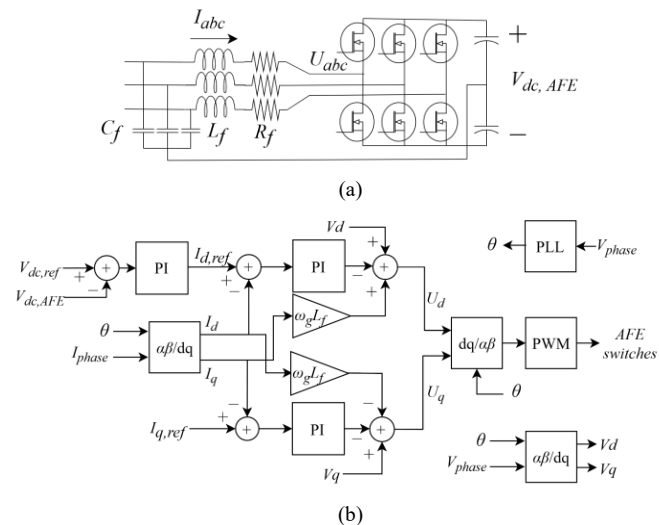

Fig. 3. AFE converter: (a) circuit diagram, (b) control block diagram.

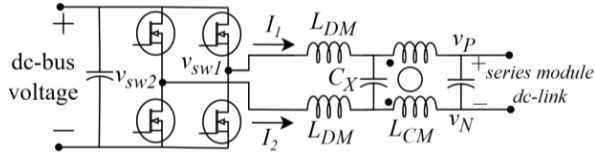

Fig. 4. Interconnecting H-bridge module circuit diagram.

Full compensation for active and reactive power leads to an operation area for dedicated active and reactive power as

$$-\frac{V_{dc}}{\sqrt{2}V_1} \le \sin(\theta_1 - \theta_2) \le \frac{V_{dc}}{\sqrt{2}V_1} \Rightarrow$$
$$-\sin^{-1}\left(\frac{V_{dc}}{\sqrt{2}V_1}\right) \le \theta_1 - \theta_2 \le \sin^{-1}\left(\frac{V_{dc}}{\sqrt{2}V_1}\right),$$
$$-\frac{V_{dc}}{\sqrt{2}} \le V_1 \cos(\theta_1 - \theta_2) - V_2 \le \frac{V_{dc}}{\sqrt{2}} \Rightarrow$$
$$-\frac{V_{dc}}{\sqrt{2}} \le V_1 - V_2 \le \frac{V_{dc}}{\sqrt{2}}. \quad (7)$$

Equation (7) sets the constraints for maximum amplitude and phase differences, which the power-flow controller can bridge for a certain module voltage level. Beyond those limits, the series-injection modules are operated in the bypass mode.

## IV. DESIGN AND CONTROL OF INTERCONNECTING H-BRIDGE MODULES

The system must be designed from aspects of common mode and differential mode currents. The interconnecting H-bridge module has two priorities: first, it must suppress the common mode current and second, it regulates the dc-link voltage. The common mode choke ($L_{CM}$) and the differential mode filter (including differential mode choke $L_{DM}$ and capacitor $C_X$) should be designed, firstly. The differential choke responsible for suppressing the differential current noise can be obtained as

$$L_{DM} = \frac{V_{dc\_bus}}{4 f_{sw} I_{ripple,DM}}, \quad (8)$$

where $f_{sw}$ and $I_{ripple}$ are respectively switching frequency and peak-to-peak desired current ripple. The capacitor $C_X$ value also can follow from

$$f_c = 10^{\frac{A}{40}} f_{sw}, \quad (9)$$

where $A$ is the required attenuation at switching frequency and $f_c$ equals to $\frac{1}{2\pi\sqrt{2L_{DM}C_X}}$. The common-mode choke size can be calculated like (8) as

$$L_{CM} = \frac{V_{dc\_bus}}{4 f_{sw} I_{ripple,CM}}, \quad (10)$$

The current ripple for the common-mode choke is much smaller than that of the differential choke due to saturation of high permeability of common-mode core.

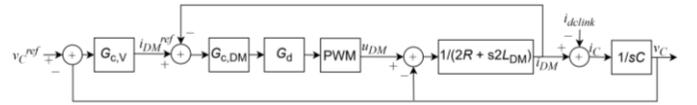

Fig. 5. Control block diagram of interconnecting H-bridge converter.

As discussed before, common-mode regulation is essential in interconnecting H-bridge modules. Considering the circuit diagram representation of the interconnecting H-bridge module, the common-mode current follows as

$$u_{CM} = \frac{R}{2} i_{CM} + \left(L_{CM} + \frac{L_{DM}}{2}\right) \frac{di_{CM}}{dt} + v_{CM},$$
$$u_{CM} = \frac{v_{sw1} + v_{sw2}}{2}, \quad (11)$$
$$v_{CM} = \frac{v_P + v_N}{2},$$

where $R$, $L_{CM}$, and $L_{DM}$ respectively are path resistance, common-mode choke inductance, and differential mode inductance. Equation (11) can be transformed into the d-q rotating frame. The corresponding d-q components are

$$\begin{aligned} u_{CM,d} &= \frac{R}{2} i_{CM,d} + \left(L_{CM} + \frac{L_{DM}}{2}\right) \frac{di_{CM,d}}{dt} \\ &\quad - \omega_g \left(L_{CM} + \frac{L_{DM}}{2}\right) i_{CM,q} \\ &\quad + v_{CM,d}, \\ u_{CM,q} &= \frac{R}{2} i_{CM,q} + \left(L_{CM} + \frac{L_{DM}}{2}\right) \frac{di_{CM,q}}{dt} \\ &\quad + \omega_g \left(L_{CM} + \frac{L_{DM}}{2}\right) i_{CM,d} \\ &\quad + v_{CM,q}. \end{aligned} \quad (12)$$

For a first-order current control loop, the decoupling terms and the output of the current controller can be written as

$$\begin{aligned} u_{CM,d} &= v_{CM,d} - \omega_g \left(L_{CM} + \frac{L_{DM}}{2}\right) i_{CM,q} + dv_d, \\ u_{CM,q} &= v_{CM,q} + \omega_g \left(L_{CM} + \frac{L_{DM}}{2}\right) i_{CM,d} + dv_q, \end{aligned} \quad (13)$$

where output current controller is

$$\begin{aligned} dv_d &= K_P \left(1 + \frac{1}{\tau_I s}\right) (i_{CM,d}^{ref} - i_{CM,d}), \\ dv_q &= K_P \left(1 + \frac{1}{\tau_I s}\right) (i_{CM,q}^{ref} - i_{CM,q}); \end{aligned} \quad (14)$$

$i_{CM,d}^{ref}$ and $i_{CM,q}^{ref}$ are usually set to zero.

Fig. 5 describes the current control loop with corresponding processing and PWM delays. The open-loop transfer function $G_{OL}$ follows

$$G_{OL} = K_P \left(1 + \frac{1}{\tau_{i,CM} s}\right) e^{-T_d s} \frac{1}{\frac{R}{2} + s\left(L_{CM} + \frac{L_{DM}}{2}\right)}, \quad (15)$$

where $T_d$ is the digital delay and equals 1.5 $T_s$. The controller must have zero steady-state error, as well as a fast transient response. Therefore, it needs to have high gain at dc and a high cross-over frequency. Since the reduced model is first-order inherently, a PI controller is sufficient. The controller bandwidth depends on the cross-over frequency ($\omega_c$) and the desired phase margin ($\varphi_m$). $\omega_c$ follows from

$$\angle G_{\text{OL}}(j\omega_c) = -\pi + \varphi_m \Rightarrow \omega_{c,\text{CM}} = \frac{\frac{\pi}{2} - \varphi_m}{T_d},$$

$$\tau_{i,\text{CM}} = \frac{L_{\text{CM}} + \frac{L_{\text{DM}}}{2}}{\frac{R}{2}}. \tag{16}$$

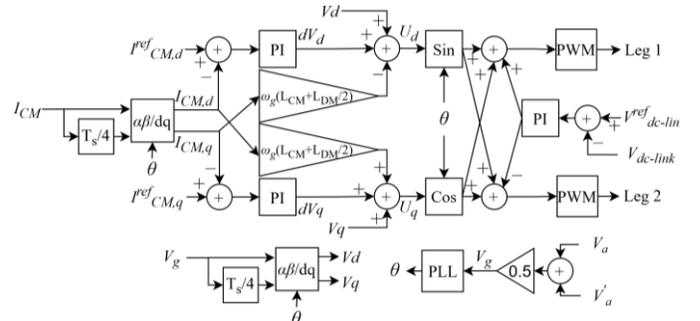

Fig. 6. Control block diagram of interconnecting H-bridge converter.

We can determine proportional gain $K_{P,\text{CM}}$ by setting the amplitude of the open-loop transfer function to unity at the cross-over frequency as

$$|G_{\text{OL}}(j\omega_c)| = 1 \Rightarrow K_{P,\text{CM}} = \omega_c \left(L_{\text{CM}} + \frac{L_{\text{DM}}}{2}\right). \tag{17}$$

Regulation of the differential component leads to the dc-link voltage. The differential components must be added to the current controller's common-mode outputs. KVL for the differential-mode components entails the differential current as

$$u_{\text{DM}} = 2Ri_{\text{DM}} + 2L_{\text{DM}}\frac{di_{\text{DM}}}{dt} + v_{\text{DM}},$$

$$u_{\text{DM}} = v_{\text{sw1}} - v_{\text{sw2}}, \tag{18}$$

$$v_{\text{DM}} = v_P - v_N = v_C.$$

Fig. 7 presents the block diagram of the desired control of differential components. It consists of inner and outer loops; the outer loop regulates the dc-link voltage, and the inner loop is responsible for the differential current. Equation (18) likewise presents a first-order equation, so the PI controller suffices for the differential components. The open-loop transfer function of the differential current can be described as

$$G_{\text{OL,DM}} = K_{P,\text{DM}}\left(1 + \frac{1}{\tau_{i,\text{DM}}s}\right)e^{-T_d s}\frac{1}{2R + s2L_{\text{DM}}}. \tag{19}$$

The procedure of tuning the proportional gain $K_{P,\text{DM}}$ and integral time constant $\tau_{i,\text{DM}}$ is similar to the procedure of the common-mode current controller. Therefore, these parameters are

$$\angle G_{\text{OL,DM}}(j\omega_c) = -\pi + \varphi_m \Rightarrow \omega_{c,\text{DM}}$$

$$= \frac{\frac{\pi}{2} - \varphi_{m,\text{DM}}}{T_d}, \tag{20}$$

$$|G_{\text{OL,DM}}(j\omega_c)| = 1 \Rightarrow K_{P,\text{DM}} = \omega_c 2L_{\text{DM}}. \tag{21}$$

The remaining parameters to be tuned are the outer-loop voltage compensator's. The open-loop transfer function of the voltage follows

$$G_{\text{OL,V}} = K_{P,V}\left(1 + \frac{1}{\tau_{I,V}s}\right)\frac{G_{\text{OL,DM}}}{1 + G_{\text{OL,DM}}}\frac{R_s}{sR_sC + 1}. \tag{22}$$

We set the proportional gain $K_{P,V}$ and integral time constant $\tau_{i,V}$ as

$$\omega_{c,v} = 0.1\omega_{c,\text{DM}} \Rightarrow \varphi_{m,v} = -\frac{\pi}{2},$$

$$|G_{\text{OL,V}}(j\omega_c)| = 1 \Rightarrow K_{P,V} = \omega_{c,V}C. \tag{23}$$

Fig. 6 presents the control bock diagram of the interconnecting H-bridge module in detail. The ac voltages at both sides of the series-injection modules are measured as reference frame to obtain the required phase angle. The error of $I_{CM}$ is compensated after transformation into the d-q frame. Then we accumulate the required deviation ($dV_d$ and $dV_q$) from the reference grid voltage ($V_{dq}$) as well as the decoupling components to the reference signal ($u_d$, $u_q$) of the PWM.

TABLE I
SYSTEM PARAMETERS VALUES

| Symbol | Description | Value |
|---|---|---|
| $V_g$ | Grid voltage | 400 V |
| $f_g$ | Grid frequency | 50 Hz |
| $Z_L$ | Line impedance | $(0.164 + 0.080j)$ Ω/km |
| $L_{\text{cable}}$ | Cable length | 1.0 km |
| $V_{\text{dc-bus}}$ | AFE's output voltage | 800 V |
| $L_F$ | AFE's filter inductor | 1.0 mH |
| $C_F$ | AFE's filter capacitor | 20 μF |
| $f_{\text{sw,H-bridge}}$ | H-bridge switching frequency | 100 kHz |
| $L_{\text{CM}}$ | Common mode choke | 2.0 mH |
| $L_{\text{DM}}$ | Differential mode choke | 100 μH |
| $f_{\text{sw,module}}$ | Series module switching frequency | 50 kHz |
| $L_s$ | Series module's inductance | 50 μH |
| $C_{\text{dc-link}}$ | Series module's dc-link capacitor | 2.2 mF |
| $f_s$ | Control sampling frequency | 10 kHz |

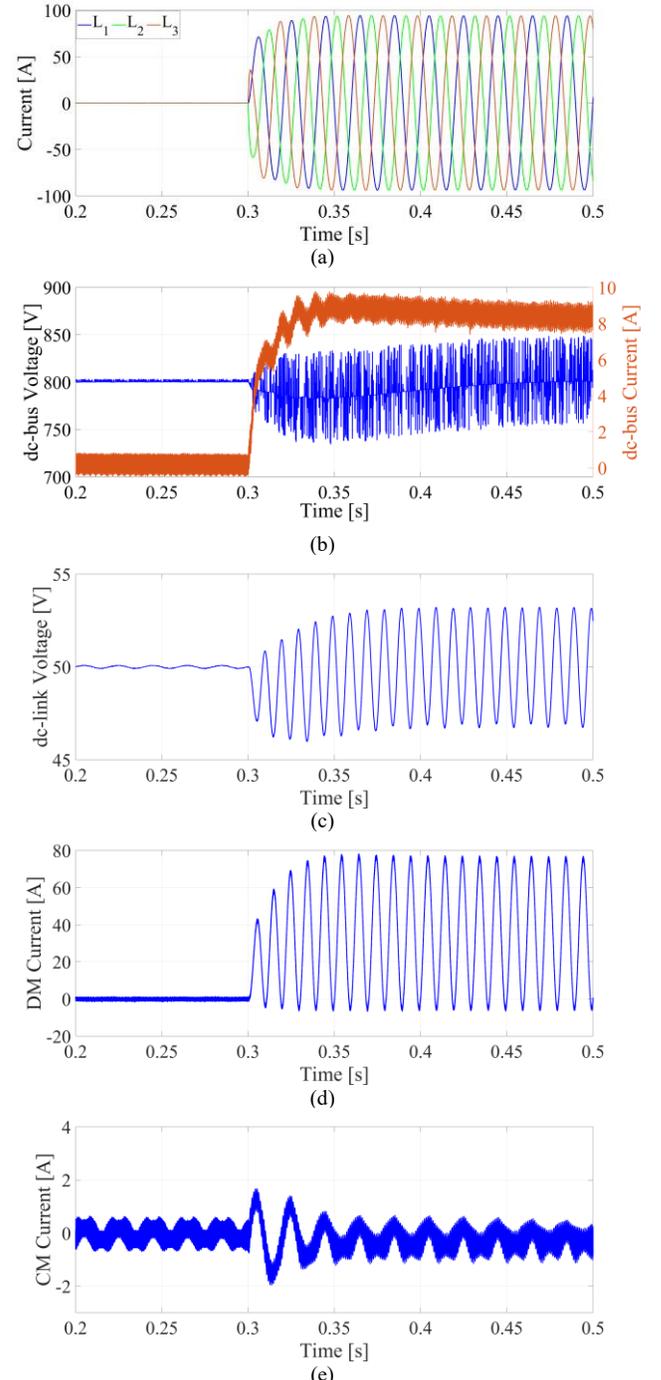

Fig. 7. Simulation results: (a) injected line current, (b) dc-bus voltage and current, (c) series-module dc-link voltage, (d) DM current, (e) CM current of interconnecting H-bridge.

## V. SIMULATION

We studied the performance of the proposed power-flow controller in MATLAB/Simulink for typical scenarios in power grids. Table I lists the system parameters.

The first scenario is to assess the functionality of the proposed power-flow controller in nominal conditions. The proposed circuit resides between two grid feeders with identical voltage vectors. Here, the maximum power injection capability depends on the impedance of the cable ($S_{max}$ = 6000 VA). Here, the controller uses $K_{P,V}$ = 12, $K_{I,V}$ = 500, $K_{P,DM}$ = 1.16, $K_{I,DM}$ = 1160, $K_{P,CM}$ = 10, and $K_{I,CM}$ = 261. Fig. 7 graphs the injected current into the feeder, dc-bus voltage/current, and series-injection module's dc-link voltage, as well as CM/DM current. From $t$ = 0.3 s on, the circuit injects a current of 95 A into the feeder. The AFE's dc-bus voltage/current and the series module's dc link are stable. So are the interconnecting H-bridge's CM/DM currents.

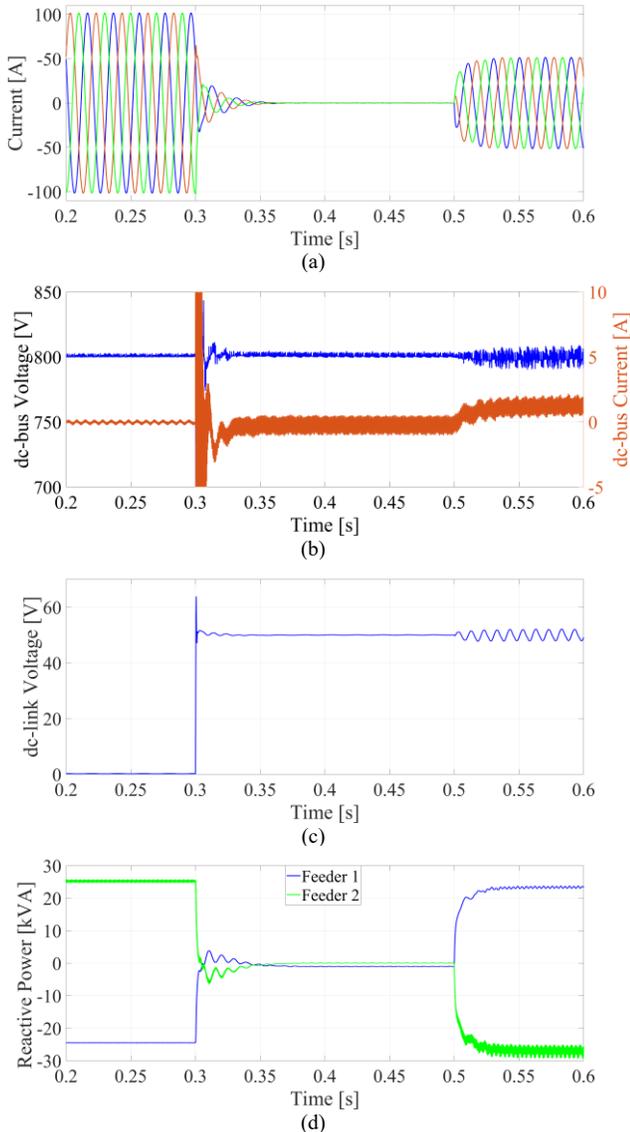

Fig. 8. Simulation results of reactive power regulation: a) line currents, (b) dc-bus voltage and current, (c) series-module dc-link voltage, (d) reactive power.

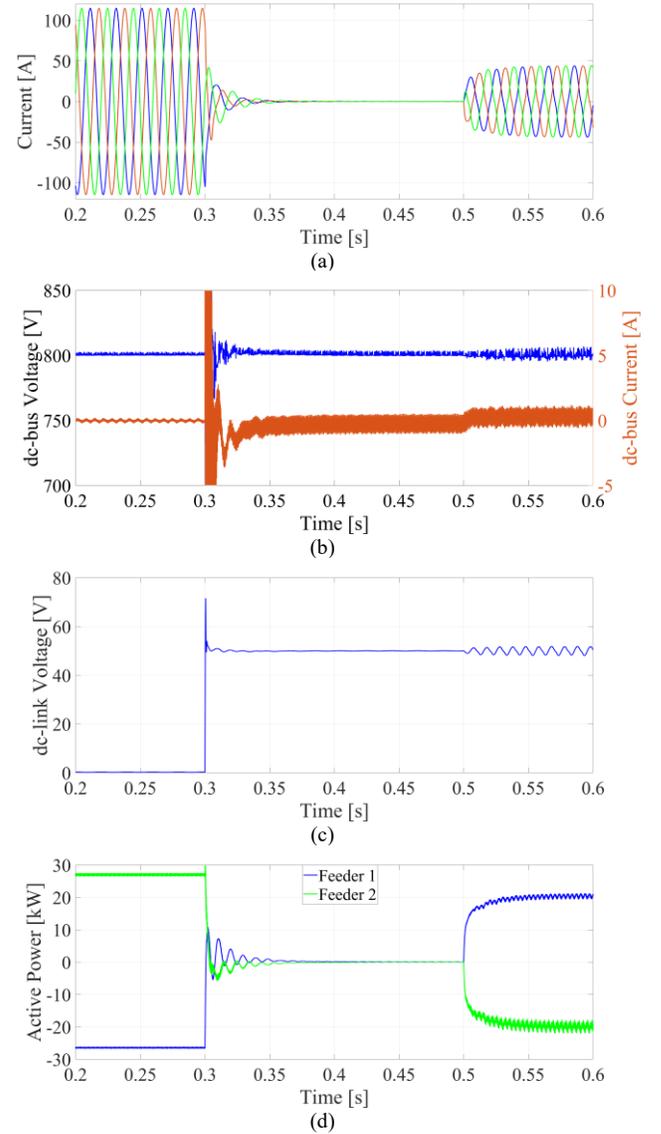

Fig. 9. Simulation results of active power regulation: (a) line currents, (b) dc-bus voltage and current, (c) series-module dc-link voltage, (d) active power.

The second scenario describes a condition in which the proposed power-flow converter connects two feeders with different voltage amplitudes. The reactive power flows from the higher voltage feeder into the lower one. The maximum voltage difference with appropriate operation of the circuit obeys (7). Fig. 8 illustrates the system variables before and after enabling the controller at $t$ = 0.3 s. Since the interconnecting H-bridges are disabled in the bypass mode of the series modules, the dc-link voltage is zero. Furthermore, the circuit not only compensates the reactive power but also injects reactive power in reverse direction at $t$ = 0.5 s.

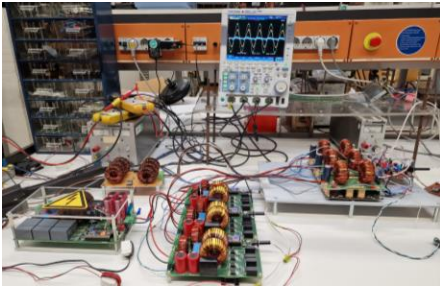

Fig. 10. Experimental setup of the proposed power-flow controller.

TABLE II
SYSTEM PARAMETERS VALUES

| Symbol | Description | Value |
|---|---|---|
| $V_g$ | Grid voltage | 110 V |
| $f_g$ | Grid frequency | 50 Hz |
| $Z_L$ | Line impedance | 0.164 + 0.080$j$ Ω/km |
| $V_{dc\text{-}bus}$ | AFE's output voltage | 400 V |
| $L_F$ | AFE's filter inductor | 700 μH |
| $C_F$ | AFE's filter capacitor | 20 μF |
| $f_{sw,H\text{-}bridge}$ | H-bridge switching frequency | 100 kHz |
| $L_{CM}$ | Common mode choke | 500 μH |
| $L_{DM}$ | Differential mode choke | 100 μH |
| $f_{sw,module}$ | Series module switching frequency | 50 kHz |
| $L_s$ | Series module's inductance | 50 μH |
| $C_{dc\text{-}link}$ | Series module's dc-link capacitor | 2.2 mF |
| $f_s$ | Control sampling frequency | 10 kHz |

In the last scenario, the circuit blocks the active power between two feeders with the maximum tolerable phase difference according to (7). Fig. 9 presents the voltages and currents as well as the active power flow before and after initiating the controller. The circuit activates at $t$ = 0.3 s to compensate the active power. Also, the active power is forced to flow in reverse direction at $t$ = 0.5 s. Thus, the proposed circuit has the capability to regulate active power in the grid lines.

## VI. EXPERIMENTS

We evaluated the proposed power-flow controller in a real system (Fig. 10, parameters per Table II). The proposed circuit is supposed to regulate power flow between two feeders independently from the impedance. A grid simulator (TI7915-350-90) feeds the power-flow controller. AFE converter, interconnecting H-bridges, and series-injection modules each have their individual controller (STM32F411). The individual controls render the system flexible and modular. We defined two scenarios of active and reactive power regulation for the circuit to evaluate its functionality.

In the first scenario, the proposed circuit injects active power into the grid in both directions (Fig. 11). The circuit can inject the active power into the grid, while reactive power is suppressed to zero.

In the second scenario, the circuit drives reactive power in both directions and regulates the active power to zero (Fig. 12) Obviously, the line current is ±π/2 in phase relative to the grid voltage.

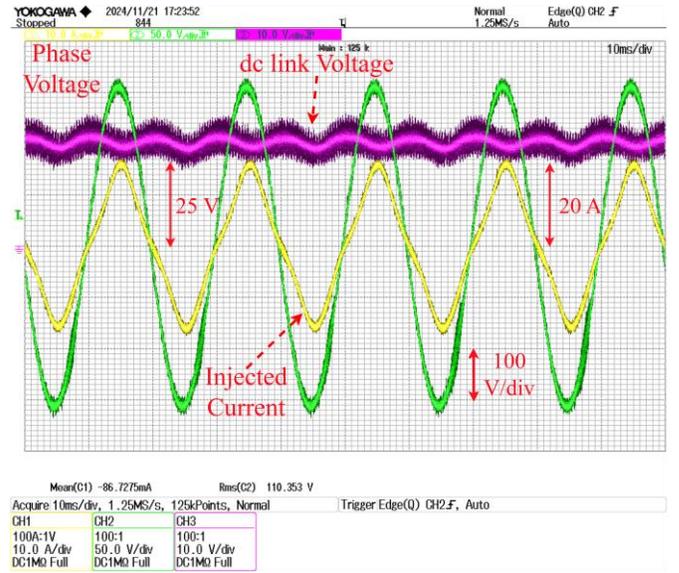

(a)

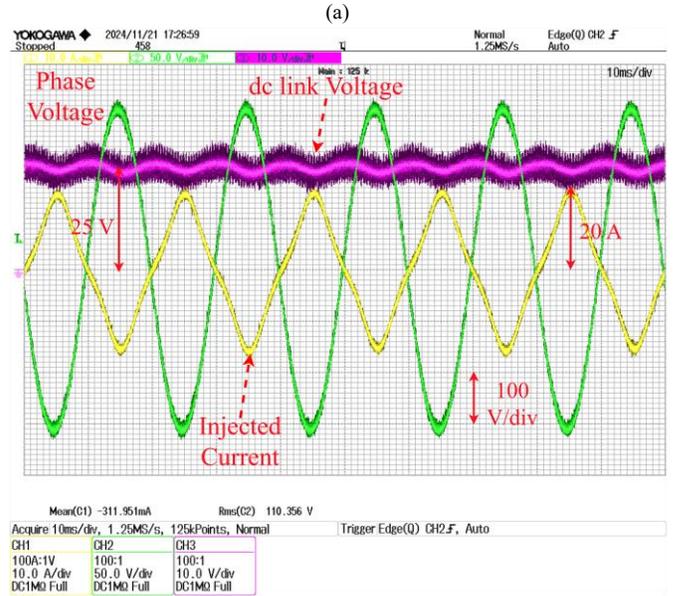

(b)

Fig. 11. Experimental results of active power regulation: (a) forward direction, (b) reverse direction.

## VII. CONCLUSION

The spread of roof-top solar systems, EV chargers, and the massive electrification of every aspect of life causes problems of stability, voltage regulation, and reverse power flow in low-voltage grids. Flexible ac transmission system (FACTS) devices such as UPFCs are well-known in high-voltage transmission grids and strongly promoted for medium-or low-voltage grids. However, their technology suffers from huge 50 Hz / 60 Hz transformers. In this paper, we presented a fully electronic power-flow controller without transformers, neither high- nor low-frequency. The elimination of transformers greatly reduces magnetics and semiconductors to drive them in comparison to other circuits.

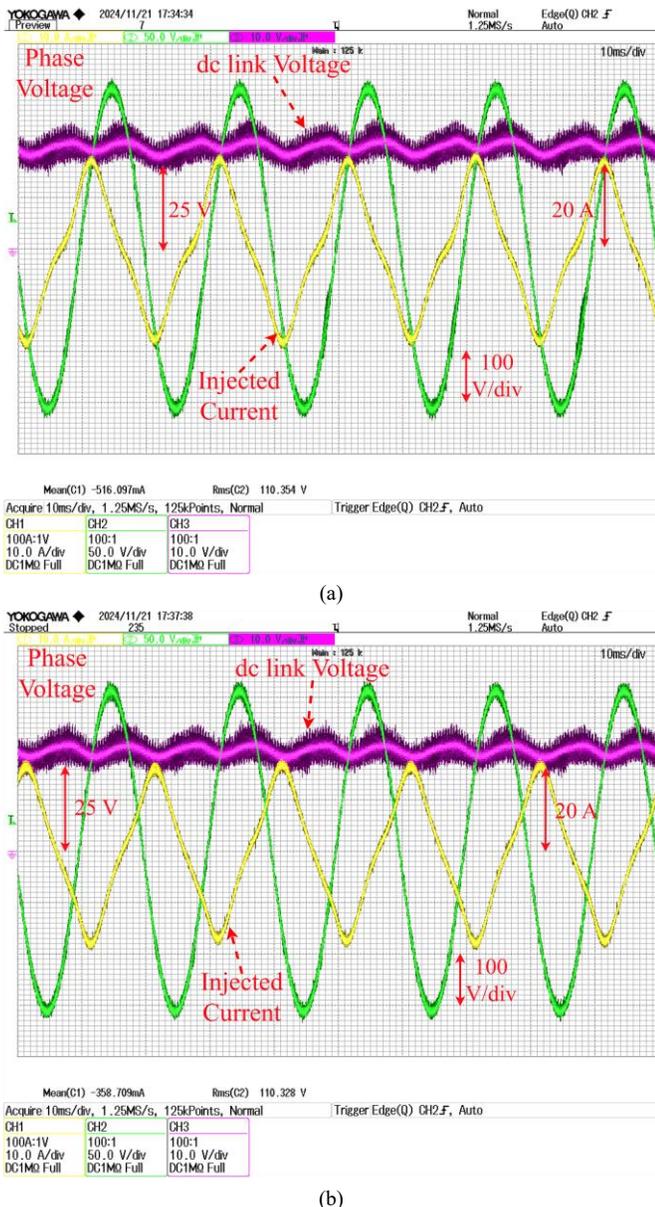

Fig. 12. Experimental results of reactive power regulation: a) forward direction b) reverse direction.